# MULTI-SENSOR INTRUSION DETECTOR SYSTEM

**BY**

**ARINDE Victor Adeshina
(183509)**

**A PROJECT REPORT SUBMITTED TO**

**DEPARTMENT OF COMPUTER ENGINEERING,
FACULTY OF ENGINEERING AND TECHNOLOGY,
LADOKE AKINTOLA UNIVERSITY OF TECHNOLOGY, OGBOMOSO,
OYO STATE. IN PARTIAL FULFILMENT OF THE REQUIREMENTS FOR THE
AWARD OF DEGREE OF BACHELOR OF TECHNOLOGY (B. TECH) IN
COMPUTER ENGINEERING**

**FEBRUARY, 2023**

# Certification

This project report with the title "**Multi-sensor Intrusion Detector System**" submitted by **ARINDE, Victor Adeshina and IDOWU Liberty Abraham** was carried out under my supervision at Ladoke Akintola University of Technology, Ogbomoso.

______________________________________          ________________
**Supervisor**                                                                                                Date
**Engr. Prof. Mrs. A. O. Oke**, B.Tech., M.Tech., Ph.D
Professor of Computer Engineering
Department of Computer Engineering
Ladoke Akintola University of Technology
Ogbomoso, Nigeria

# Attestation

I hereby attest that this research was carried out in the Department of Computer Engineering, Faculty of Engineering and Technology, Ladoke Akintola University of Technology, Ogbomoso, Nigeria.

______________________________________                                              ________________
**Head of Department**                                                                                                        Date
**Engr. Prof. J. B. Oladosu**, B.Tech., M.Sc., Ph.D
Professor of Computer Engineering
Department of Computer Engineering
Ladoke Akintola University of Technology
Ogbomoso, Nigeria

## Acknowledgement

My heartfelt appreciation goes to God Almighty, the one who saw me through in the pursuit of my undergraduate programme in Lautech. The faithful Father who protected me and supplied all my needs. Thank you lord! My sincere gratitude goes to my project supervisor, Engr. Prof. Mrs. A. O. Oke, for her wonderful supervision and support that made this project work a successful one; and for teaching me to give attention to the details of my work. Thank you so much ma and may God bless you and your family (Amen).

**Abstract**

Security, defined as protection against external threats, is a critical concern for homes and offices. Intrusion, characterized by unauthorized access, presents a significant challenge to maintaining security. This research aims to address this issue by designing and implementing an automated intrusion detection system utilizing a combination of sensors and communication technologies.

The research introduced an automated intrusion detection system for homes and offices, combining sensors such as a PIR sensor for detecting unauthorized motion, magnetic switches for unauthorized entry detection, and a GSM module for notifying property owners. Employing the ATmega328P microcontroller, sensor data is analysed to generate early intrusion alerts, prompting phone call notifications via the GSM module. Practical implementation involved breadboarding, soldering, and rigorous testing, ensuring proper functionality under real-world conditions.

The implemented intrusion detection system effectively utilizes magnetic switches and a Passive Infrared (PIR) sensor to detect unauthorized entry and motion within the premises, respectively. Upon detection, the system promptly analyses the situation and alerts the property owner via phone call, enabling swift response measures. This real-time notification system enhances proactive security management, minimizing the risk of further intrusion and ensuring the safety of the property.

The multi-sensor intrusion detection system, incorporating PIR sensors, magnetic switches, and a GSM-based phone call gateway, effectively alerts property owners of unauthorized intrusions in real-time. Demonstrating its efficacy through rigorous testing, the system offers enhanced security for both residential and commercial environments.

# TABLE OF CONTENTS

# LIST OF FIGURES

# CHAPTER ONE

# INTRODUCTION

## 1.1 Background of Study

Security is defined as the protection from action from without or subversion from within. It is the degree of protection against danger, loss and criminals. A security system is literally a means or method by which something is secured through a system of interworking components and devices (Ahmad *et al*, 2019). Intrusion is a security event, or a combination of multiple security events, that constitutes a security incident in which an intruder gains, or attempts to gain, access to a system (e.g., homes, offices, network and processes) or system resource without having authorization to do so. It is said to be any kind of unauthorized activity that causes interference; and as it is unwanted or unauthorized, it is normally and mostly with bad intentions (Rao *et al*, 2014; Khraisat *et al*, 2019).

Increase in crimes, abduction, and robbery in today's world has led to advancement in home monitoring and alarm security systems. Most of these crimes happen due to the absence of the house or office owner and at times while the house occupants are present but are either asleep or carried away by other activities. Besides, having only elderly people in the house also does not prevent homes from being safe from robbery (Mustafa *et al*, 2020). Homes, industries, schools, and organizations today are invaded mostly by force either through breaching of windows, entering through a cutting ceiling or even entering through a closed door or sometimes even an open window (Olarewaju *et al*, 2017).

Tasie *et al*, (2020) opined that the alarming rate of insecurity in Nigeria and indeed other countries all over the world in the last few decades facilitates the need to have home protection gadgets that will alert home owners against intruders. Traditionally, security guards who manually provide surveillance during day and night guarded homes, but it was not fool-proof as it was only normal for them to have momentary lapse in concentration. Systems that automatically detect intrusion in homes and offices were found to be feasible solution because they provide prompt notification to the property owner in real time and furthermore help to apprehend the intruder. An automatic intrusion detection system essentially consists of CCTV/webcam, alarms, SMS and various sensors.

The fact that security alarm systems are present in an environment or house frequently serves as a deterrent to terrify an intruder before a forceful attempt inside, therefore having them in homes or places of business will boost user comfort and sense of security. (Adeline *et al*, 2017). These devices function as inputs that trigger the security alarm. Some of the security alarm sensor technologies that have been established over the centuries are passive infrared sensor (PIR), sound sensor, and magnetic switches.

Passive Infrared Sensor (PIR) which is an electronic sensor used in motion detectors such as automatically triggered lighting devices and protection systems that measure devices emitting infrared light in their field of view. Each body with a temperature above zero releases heat energy, which is in the form of radiation. PIR sensors detect infrared radiation that is reflected or released from the target instead of measuring or sensing heat (Himadri *et al,* 2022). The sound sensor module is a low-cost electronic sensor that can easily detect sound. It is generally used for detecting sound intensity. This sensor has a microphone for detecting sound intensity. The electric microphone detects the sound wave and then sends it to the sensor circuit board which consists of a voltage comparator IC LM393 and potentiometer. Comparator IC LM393 processes this signal and converts it into a Digital Output. The potentiometer is used to adjust the sensitivity of the sensor (EletroDuino 2020).

Magnetic Switches is most usually used in electromechanical appliances that trigger when the magnet and contact are alienated. It is mainly used on windows or doors; these switches are the prevailing detecting gadgets in detecting closing or opening of windows or doors. The detectors are reliable and cheap. This type of sensors usually comprises two sections, a contact that is usually installed on the window or doorframe and an activating magnet that is mounted on the door (Muhammad *et al,* 2019).

The busy lifestyle of people is leading to the necessity of controlling the devices at home remotely. Currently, calls are not the only thing made on mobile phones. Since calls may be used for a variety of things, mobile phones are changing how they are used as technology advances. Instead of merely being used as phones, they can be utilized as clocks, calendars, or remote controls. Today, smart phones are available in the market with different applications and hardware which can be implemented without any further development or enhancement. With the help of the GSM network, a mobile can be used to implement a smart home by controlling devices and getting phone calls on robbery (Nwalozie *et al*, 2015).

Provisioning an automated intrusion detection system will help to reduce break-ins into homes and offices. Hence, this work designed and constructed an automated intrusion detection system, using the combination of Microcontroller ATMega328, PIR motion sensor, magnetic switches and GSM Module.

## 1.2 Statement of the Problem

The rate of Crime in the world is increasing day by day due to urbanization, unemployment, poverty, economic recession, and social inequality, which will bring chaos to the country. Most of the crimes that are usually done are abduction, robbery, theft and house breaks, but the most common one done today is armed robbery (Muhammad *et al.,* 2019). Security is the degree of protection against danger, loss and criminals.

Over the years, different researchers have developed several smart devices for intrusion detectors for home and office use that involved the process of using GSM/SMS and Wi-Fi based technology methods, and each of these researchers accomplished success and found a solution to an issue. (Ayush and Joshi, 2012; Nwalozie *et al.,* 2015; Arun and Saravana, 2021; Yuda *et al.,* 2021; Usiade and Adeoye, 2022). Nevertheless, most of these researchers did not work for a means of reducing cost by using a device notifying the home or office owner via phone call. Therefore, in order to reduce cost, this research will use phone calls which is cheaper than SMS messages to deliver instant alerts. This initiative will guarantee that users can react quickly to any dangers in addition to making intruder detection systems affordable.

## 1.3     Aim and Objectives

The Aim of this project is to design and construct a multi-sensor intrusion detector system. The objectives are to:

1. Design a multi-sensor intrusion detector system.

2. Construct a model for the system.

3. Implement the designed system.

## 1.4     Significance of Study

Intruder detectors can be integrated with other security measures like alarms and surveillance cameras in addition to detecting invasions. If a crime does occur, this integration may aid in capturing the necessary evidence. Office and home owners can rest easier knowing that dependable intruder detection systems are installed on buildings. The result may be an increase in general wellbeing. In the event of an intrusion, intruder detectors smart devices can trigger automatic alerts via phone call to the home and office owner, leading to quicker response times and potentially preventing further damage or theft.

Intruder detection systems can be designed to suit the specific needs of a home or office. This customization ensures that the system is effective and doesn't lead to false alarms, which can be a common issue with generic systems. Designing intruder detectors involves incorporating cutting-edge technology such as motion sensors, video analytics, and artificial intelligence. Advancements in these technologies can lead to more accurate and efficient intrusion detection.

## 1.5    Scope of Study

This study is organized towards the design and construction of an intruder detector system for home and office use which will detect the presence of an intruder and give rapid response to the owner via phone call by reducing SMS alert fee. Different sensors: microcontroller, GSM module, PIR sensor, Magnetic switches will be connected together to detect unwanted presence and movement and closing or opening of windows or doors in a home and office in the absence of the owner.

# CHAPTER TWO

# LITERATURE REVIEW

## 2.1 Intrusion Detection System

Growing focus on the perimeter security of national properties, both at home and abroad, has facilitated the development of technologies capable of detecting possible threats, such as people or vehicles approaching them (Aniket *et al,* 2021). Therefore, intrusion detection is the process of monitoring the events occurring in a system or network and analyzing them for signs of intrusions, defined as attempts to compromise the confidentiality, integrity, availability, or to bypass the security mechanisms of a system or network. Intrusions are caused by attackers accessing the systems from the Internet, authorized users of the systems who attempt to gain additional privileges for which they are not authorized, and authorized users who misuse the privileges given them (Bace *et al*, 2000). An Intrusion Detection System (IDS) is a software or hardware tool used to detect unauthorized access of a system or network (Prakash*,* 2013). The main aim of intrusion detection is to monitor network assets to detect anomalous behavior and misuse in network (Rajasekaran*,* 2013).

Ankiet *et al,* (2021) suggested that Intrusion Detection Systems (IDSs) use a wide range of sensors to identify unauthorized attempts to enter secure arenas and to provide security response teams with warning signals as well. Passive Infrared (PIR) sensors, proximity sensors, microwave sensors, video detectors and magnetic switches are the major sensor technologies for interior safety. By embedding electronic hardware with these sensors, software, and networking gear into any entity within the physical world, individuals can now track house/office/store security conditions on a continuous basis; with the recent increase in availability and usage of smart IoT devices and the ubiquity of smartphones.

## 2.2　Importance of Home Security Alarm System

Muhammad *et al,* (2019), opined that the rate of Crime in the world is increasing day by day due to urbanization, un employment, poverty, economic recession, and social inequality, which will bring chaos to the country. Most of the crimes that are usually done are abduction, robbery, theft and housebreaks, but the most common one done today is armed robbery. This disturbing increase rate of crime in the world today, thus, threatened the life and properties of the people. A security alarm system should be installed as a standard device in homes or the environments needed to be secure.

Therefore, a security alarm system is an essential device in protecting organizations, industries or buildings and improving the quality of people's life since is going to be an actual means of decreasing the threat of abduction, burglary, and thefts in the world today. Therefore, the importance of installing the security alarm system are to protect valuable belongings in the home. An alarm system in the household will scare off burglars and it will also send notification to local authorities if someone did try to break in to the property.

With the help of the installation of the security alarm system installed, indoor and outdoor cameras can be remotely monitored i.e. the house and what's happening around the environment or surroundings and can be watched. With the help of the security alarm system installed, it may prevent the loss of valuables and properties which can lead to gigantic financial damage. The installation of the security alarm system may provide the homeowners to be secure or with self-confidence and relaxation of the mind that their properties are secured.

Some of the disadvantages of a home security alarm system include vulnerability, when users of security cameras, try to stay updated on the latest in security systems, intruders and criminals are also doing the same too. A clever trespasser will probably know all about them and may have figured out a way to go undetected. Further, tech-savvy criminals might have understood the technology and worked out ways to disable/disconnect them from their power source. Privacy and cost are also part of the disadvantages.

**2.3    Microcontroller Unit**

Microcontroller can be considered as a self-contained system with a processor, memory and peripherals and can be used with an embedded system. Microcontrollers are designed for small or dedicated applications and are often used in automatically controlled products and devices, such as automobile engine control systems, implantable medical devices, remote controls, office machines, appliances, power tools, and toys (Oke *et al,* 2015). A microcontroller is said to be a miniaturized computer in a chip of silicon and can accept instructions and follow those instructions (Izang *et al,* 2018) The operation of a microcontroller can be changed or modified by writing a new set of instructions or program in assembly language (Oke *et al,* 2015).

ATMega328 microcontroller is a very versatile device that is adequate for the role intended in fire outbreak detection systems. It can monitor and control the rest of the hardware and also provide 40mA of drive current for any device connected to its ports. The board also has an on board 5volt regulator, which means it can run from a power supply even higher than 5 volts (Izang *et al,* 2018). By reducing the size and cost compared to a design that uses a separate microprocessor, memory, and input/output devices, microcontrollers make it economical to digitally control even more devices and processes (Falohun *et al,* 2016).

ATmega328 is basically an Advanced Virtual RISC (AVR) micro-controller. It supports the data up to eight (8) bits. ATmega-328 has 32KB internal built-in memory. ATmega328 has IKB Electrically Erasable Programmable Read Only Memory (EEPROM). This property shows if the electric supply supplied to the micro-controller is removed, even then it can store the data and can provide results after providing it with the electric supply. Moreover, ATmega-328 has 2KB Static Random-Access Memory (SRAM). It has 8 Pin for ADC operations, 3 built-in Timers, two of them are 8 Bit timers while the third one is 16-Bit Timer. The Atmega328 IC has an operating range from 3.3V to 5.5V but normally we use 5V as a standard. Its excellent features include the cost efficiency, low power dissipation, programming lock for security purposes, real timer counter with separate oscillator. In this work, "microcontroller" will be used interchangeably except explicitly stated. The ATMega382P IC and pin out are shown in Figure 2.1.

Another microcontroller option worth considering is the STM32 series, which offers enhanced performance and versatility for embedded system applications. While the ATMega328 excels in various small-scale applications, the STM32 microcontrollers, developed by STMicroelectronics, provide a higher level of computational power and advanced features. Featuring 32-bit ARM Cortex-M cores, the STM32 devices outperform the ATMega328 in processing capabilities, enabling them to handle more complex tasks efficiently. Moreover, STM32 microcontrollers integrate a wide range of peripherals, including advanced ADCs, DACs, timers, and communication interfaces, facilitating the implementation of sophisticated features in embedded systems. With comprehensive memory configurations and support for external memory interfaces, STM32 devices offer greater flexibility in memory management compared to the ATMega328 (Izang *et al,* 2018).

Additionally, the STM32 family benefits from a robust development ecosystem, comprising integrated development environments, software libraries, and extensive documentation, ensuring seamless development and deployment processes. By considering alternatives like the STM32 alongside the ATMega328, developers can evaluate trade-offs in performance, feature set, and cost to select the most suitable microcontroller for their specific application requirements.

ESP32 is another type of microcontroller that is equipped with a dual-core processor and built-in Wi-Fi and Bluetooth connectivity, making it particularly suitable for wireless and IoT applications. Developed by Espressif Systems, the ESP32 offers enhanced processing power and versatility, enabling simultaneous execution of tasks and seamless communication with other connected devices or networks. With its integrated Wi-Fi and Bluetooth capabilities, the ESP32 simplifies the design of IoT devices and facilitates real-time data transmission. Moreover, the ESP32 features a rich set of peripherals, including ADCs, DACs, PWM channels, SPI, and I2C interfaces, providing extensive flexibility for hardware integration and control. Its advanced power management features further optimize energy efficiency, making it ideal for battery-operated applications. Supported by a vibrant development community and a wide range of development tools and resources, the ESP32 offers a compelling alternative to the ATMega328 for projects requiring wireless connectivity and advanced features. By exploring options like the ESP32 alongside the ATMega328, developers can tailor their choice of microcontroller to meet specific project requirements, whether it be for traditional embedded systems or cutting-edge IoT applications (Yuvaraj and Ramesh, 2012).

**2.4   GSM MODULE**

This a device that uses GSM mobile phone technology to provide a wireless data link to a network GSM modems are used in mobile telephones and other equipment that communicates with mobile telephone networks and they accept SIM card to identify a device to the network (Abhishek, 2015). SIM800L is a dual-band GSM/GPRS engine that works on frequencies EGSM

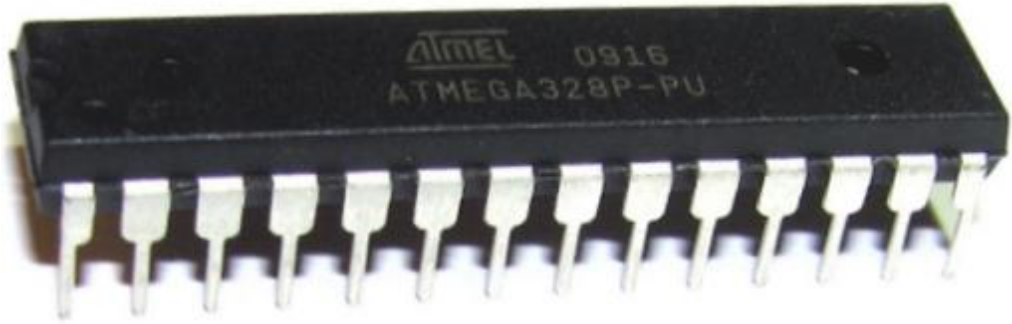

Figure 2.1a: ATMega328 IC

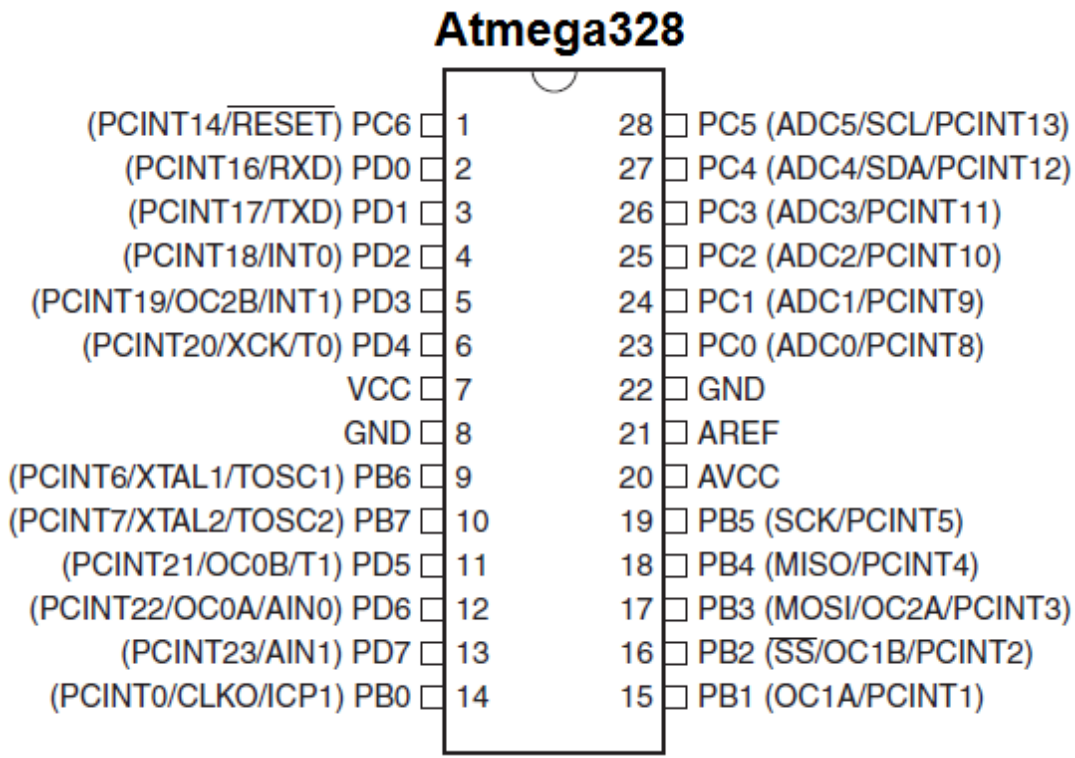

Figure 2.1b: ATMega328 IC Pinout

Atmega328 IC and Pinout, 2018

800MHz and DCS 1800MHz SIMBOGA features GPRS multi-slot class 10/ class 8 (optional) and supports the GPRS coding schemes CS-1, CS-2, CS-3 and CS-4.

SIM800L is readily available GSM/GPRS module, used in many mobile phones, PDA, for developing IOT (Internet of Things) and Embedded Applications. The communication with this module is done through UART or RS232 Interface. The data is sent to the module or received from the module though UART interface. The module is typically connected to +40V standard power supply, it can work on +4 SV regulated power and any higher voltage may damage the module. GSM modules operates over a subscription to a mobile operator, just like a mobile phone. While these GSM modems are most frequently used to provide mobile internet connectivity, many of them can also be used for sending and receiving SMS and MMS messages. A GSM modem can be a dedicated modem device with a serial, USB or Bluetooth connection, or it can be a mobile phone that GSM modem capabilities. The SIM800L GSM Module and Sim800 pin out is shown in Figure 2.2a. and Figure 2.2b respectively.

Another viable option worth considering for wireless communication in embedded systems is the Quectel M95 module. Similar to the SIM800L, the Quectel M95 is a GSM/GPRS module designed to provide wireless connectivity to mobile networks. It operates on the GSM/GPRS 850/900/1800/1900 MHz frequency bands, offering compatibility with a wide range of networks worldwide. The Quectel M95 module features a compact design and supports GPRS multi-slot class 12/10 for efficient data transmission. It also offers robust communication capabilities, supporting various data coding schemes and protocols to ensure reliable data transfer over GSM networks (Abhishek, 2015). Like the SIM800L, the Quectel M95 module interfaces with microcontrollers or other devices via UART, enabling seamless integration into embedded systems. It can be powered by a standard +5V power supply, making it compatible with a wide range of applications.

## 2.5 Types of Sensors

There are different types of sensors that are commonly used in various applications. All these sensors are used for measuring one of the physical properties like Temperature, Resistance, Capacitance, Conduction, Heat Transfer etc. some of the types of sensors include temperature sensor, proximity sensor, accelerometer, Infrared (IR) sensor, pressure sensor, light sensor, ultrasonic sensor, touch sensor, humidity sensor, position sensor etc.

### 2.5.1 Passive Infrared sensor or Pyroelectric Infrared Radial sensor (PIRs)

Pyroelectric Infrared (PIR) sensor is a device that can sense the infrared (IR) light within its viewing range. This sensor is a passive device that simply measures the changes in the IR levels emitted by surrounding objects. Since this device is a passive measuring device it is also called "Passive Infrared" sensor. PIR will detect any object emitting IR radiation, heat or changes in the background IR level. IR radiating objects include humans, animals, vehicles and wind (Emin, 2009). The PIR sensor is the core part of the system. The system basically function based on infrared radiation, which is emitted from human body (Herbert, 2012). Therefore, PIR sensor is widely used in security system to detect the motion of human.

Infrared light (IR) is the electromagnetic radiation with a wavelength between 0.7 and 300 micro meters. Human beings are the source of infrared radiation. It was found that the normal human body temperature radiate IR at wave lengths of 10 micro meters to 12 micrometers (Yuvaraj

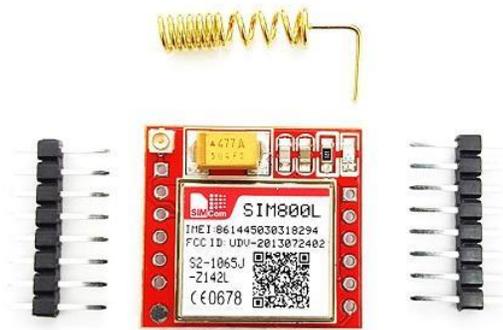

Figure 2.2a: SIM800L GSM Module

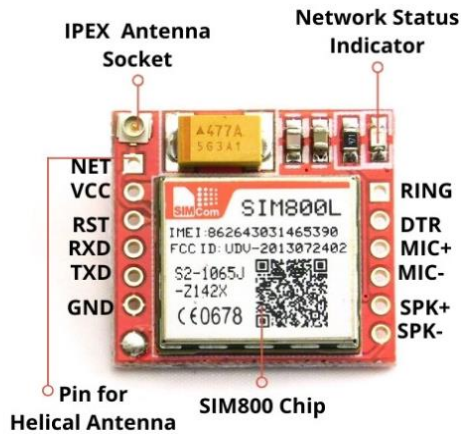

Figure 2.2b: SIM800L Chip

SIM800L GSM Module and Chip

(Electronic Projects, Embedded News and Online Community - Electronics-Lab, n.d.)

and Ramesh, 2012). PIR sensors are passive electronic devices which detect motion by sensing infrared fluctuations (David and West, 2012). It has three pins (gate, drain and source). After it has detected IR radiation difference, a high is sent to the signal pin. PIR sensor is made up of crystalline material that generates a surface electric charge when exposed to heat in the form of IR (Yuvaraj and Ramesh, 2012). This change in radiation striking the crystalline surface gives to change in charge. The sensor elements are sensitive to radiation of wide range but due to the use of filter window that limits the sensitiveness to the range 8 to 14 micrometers which is most suitable to human body radiation. The PIRs chip is shown in Figure 2.3.

### 2.5.2 Magnetic Switches

An electrical switch that is used to make or break contact within a magnetic field is known as a magnetic switch. A magnetic switch (also known as a reed switch) is composed of two partly overlapping ferromagnetic blades. It is a component of a low-cost alarm device. One part of the device contains the switch and the other a magnet. The distance between the parts determines the state of the magnetic switch (Sergej, 2020). The magnetic switch function is to stay activated if a strong magnetic field is there sufficiently and deactivated once the magnetic field is removed. Magnet switch is used where moving elements is not possible or not desirable to make direct contact through the switch like in unstable environments, submerged within liquids, etc. These switches are enclosed in glass material for maintaining the integrity of the magnetic field and components. So, this material also protects the switch from external conditions. When selecting or designing a switch, it is necessary to consider these parameters like type of application, power requirements, circuitry magnetic sensitivity, and operating environment. The Magnetic switch shown in Figure 2.4.

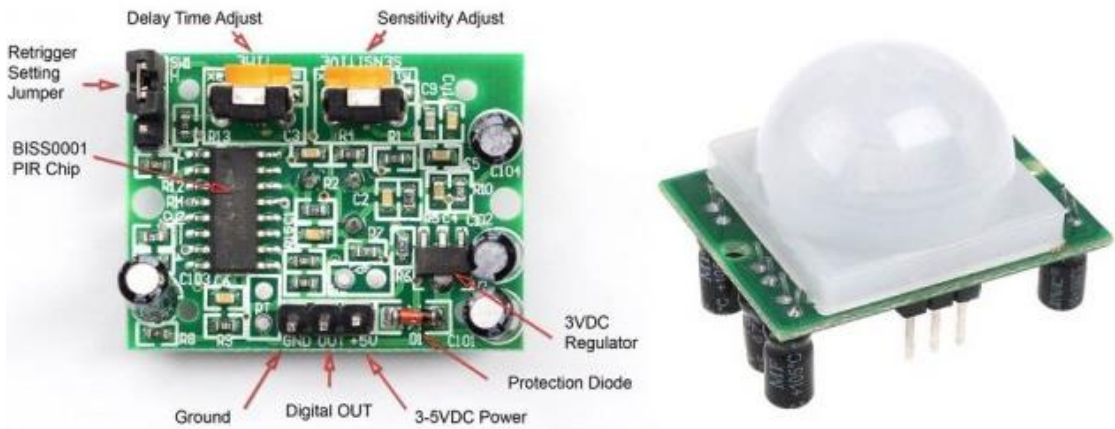

Figure 2.3: PIR Sensor chip

(Electronic Projects, Embedded News and Online Community - Electronics-Lab, n.d.)

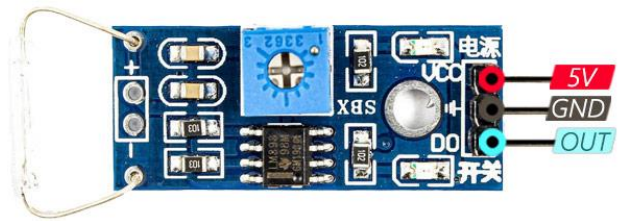

Figure 2.4: Magnetic Switches

(Electronic Projects, Embedded News and Online Community - Electronics-Lab, n.d.)

### 2.5.3 Sound Sensor

The sound sensor is one type of module used to notice the sound. Generally, this module is used to detect the intensity of sound. The applications of this module mainly include switch, security, as well as monitoring. The accuracy of this sensor can be changed for the ease of usage. This sensor employs a microphone to provide input to buffer, peak detector and an amplifier. This sensor notices a sound, and processes an output voltage signal to a microcontroller. After that, it executes required processing. This sensor is capable to determine noise levels within DB's or decibels at 3 kHz 6 kHz frequencies approximately wherever the human ear is sensitive. The working principle of this sensor is related to human ears. Because human eye includes a diaphragm and the main function of this diaphragm is, it uses the vibrations and changes into signals. Whereas in this sensor, it uses a microphone and the main function of this is, it uses the vibrations and changes into current otherwise voltage (Scribd, 2007). The Sound Sensor Module is shown in Figure 2.5.

### 2.5.4 Temperature Sensor

Temperature sensors are essential components used in a wide range of applications for measuring the thermal conditions of the surrounding environment or specific objects. They operate based on various principles, including resistance temperature detectors (RTDs), thermocouples, and thermistors. RTDs utilize the change in electrical resistance of metal alloys with temperature variations, providing accurate and linear temperature measurement over a wide range. Thermocouples, on the other hand, generate a voltage proportional to the temperature difference between two junctions of dissimilar metals, making them suitable for high-temperature environments and rugged applications. Thermistors are semiconductor devices whose electrical resistance changes nonlinearly with temperature, offering high sensitivity and fast response times,

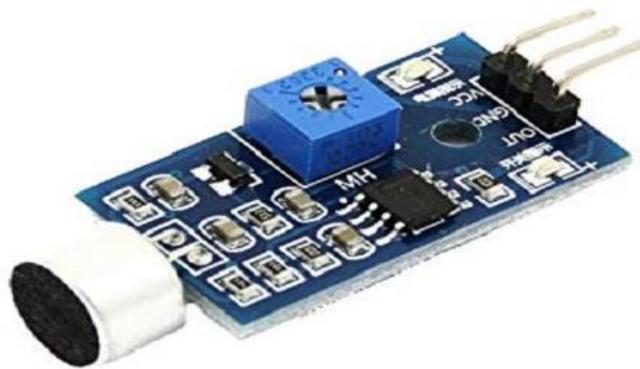

Figure 2.5: Sound Sensor Module

(Electronic Projects, Embedded News and Online Community - Electronics-Lab, n.d.)

ideal for precise temperature control in electronic circuits and medical devices (Scribd, 2007). The Temperature Sensor Module is shown in Figure 2.6.

**2.5.5   Proximity Sensor**

Proximity sensors are non-contact devices that detect the presence or absence of nearby objects within a certain range. They utilize various technologies, including inductive, capacitive, and ultrasonic sensing principles, to detect objects based on their physical properties. Inductive proximity sensors generate electromagnetic fields that change when metallic objects enter their detection range, making them suitable for detecting metal objects in industrial automation and machinery. Capacitive proximity sensors detect changes in capacitance caused by the presence of objects with different dielectric constants, making them ideal for detecting non-metallic objects such as plastics and liquids. Ultrasonic proximity sensors emit ultrasonic waves and measure the time it takes for the waves to reflect back, enabling accurate distance measurement and object detection in various applications (Scribd, 2007). The Proximity sensor is shown in Figure 2.7.

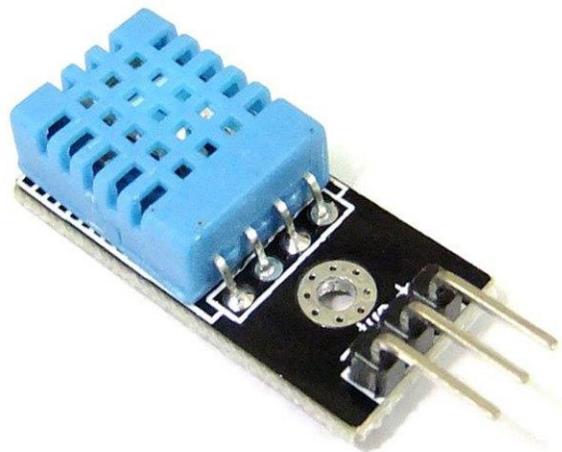

Figure 2.6: Temperature Sensor Module

(Electronic Projects, Embedded News and Online Community - Electronics-Lab, n.d.)

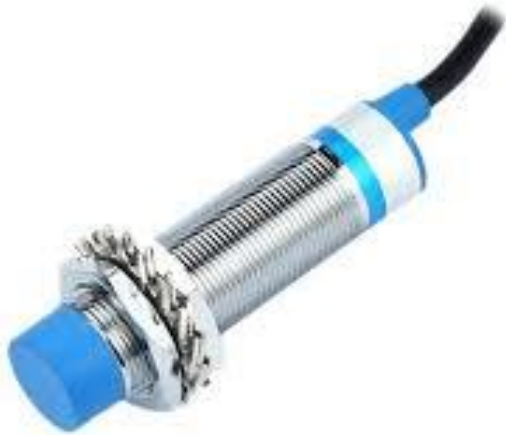

Figure 2.7: Proximity Sensor

(Electronic Projects, Embedded News and Online Community - Electronics-Lab, n.d.)

## 2.6 Related Works

Smart home security system has been a feature of science fiction writing for many years, but has become practical since the early 20th century following the widespread introduction of modern internet services into the home and rapid advancement of information technology (Usiade and Adeoye, 2022). Growing security concerns have become more noticeable in daily life, particularly when individuals are away from their residences (Azid and Kumar, 2011). To address these issues, modern technology has been harnessed to develop solutions for safeguarding us from various hazards. The field of security research has seen significant progress, as evidenced by recent studies in this area.

Ayush and Joshi, (2012) developed a low-cost GSM/GPRS based wireless home security system which includes wireless security sensor nodes and a GSM/GPRS gateway. It has the following features: (a) low cost, (b) low power consumption, (c) simple installation, (d) fast response and (e) simple user interface. In general, GSM modem acts as the interface between the users and the sensors nodes. There are 3 types of sensor nodes applied in the system which include the door security nodes, infrared sensor nodes, and fire alarm nodes. The architecture includes components like; filters, amplifiers, analog to digital converters and communication interfaces. The system used a wireless transceiver module to transfer data between gateway and sensor nodes. Every sensor node comprises a microprocessor and a wireless transceiver module. The function of the microprocessor is to receive and analyze the signal from the sensors' node as well as the current status of the nodes. This system also consists of a sleep timer and switch mode pump circuit, which reduces of the power consumption.

Nwalozie *et al.,* (2015) implemented a smart home system by controlling the electronic devices at home remotely with the help of a mobile device and getting alerts on intrusion or movement around the restricted premises. SIM900-GPRS module and the microcontroller are used to communicate between the mobile phone and the devices and sensors installed at home. The mobile phone is used as a controller from anywhere in the world if the GSM network is available. In addition, three sensors were used as a heat detector, motion detector and intrusion detector which trigger the alarm upon reaching the critical limit. The system is limited to the area with the GSM network available and the whole system does not work without the network.

Arun and Saravana, (2021) developed "Security and Power Saving System Using Embedded Technology" to explained how the PIR detector is employed in safety and power saving system. The PIR detector unendingly monitors the human motion in wherever it is placed. As an influence system, the detector senses the human motion then the corresponding fan and light that are connected to the microcontroller became on. If the sensors are not detecting any human motion then it becomes off. It controls the required fans and lights in keeping with the persons in an exceedingly space and it saves the current or power employed in homes, offices, schools and industries etc. As a security system, it's used as human unwelcome person system by detective work unauthorized access, temperature increment; IR detection monitored the standing of every individual detector. It defended homes or offices with the protection alarm. The PIR detector cowl a distance of regarding twenty feet. So, among this vary, it ready to discover any human motions. The distinctiveness of the research is not solely alerting the neighbors by alarm, it conjointly sends SMS to a specific mobile number that is already programmed into the system.

Usiade and Adeoye., (2022) work presented the design and construction of a smart home security and intrusion detection system using gas sensor, fire/smoke sensor, motion sensor, GSM module, buzzer and relay. After the construction and component assembly, it was tested that they were responding to the GSM modem as detected by the infra-red sensors, high temperature sensor and gas sensor etc. But misuse of the system by end users may probably lead to lapses in the system performance. The system was designed and constructed in such a way that maintenance and repairs are easily done in the faults. The design and construction of the GSM based intelligent home security system involved researches in different aspects of physics/electronics technology; this include; power electronics, operational amplifier, telecommunication, and software engineering. When the PIR finds intruders (in form of variation in temperature, gas leakage, pressure, etc.), the relevant sensing device(s) respond and the microcontroller sends encoded alarm signal to the wireless sensor network established in home. The moment the alarm signal is received, it will send alarm short message to the users (owners of the building) through GSM network immediately. The design analysis and calculations were carried out and finally, a positive result was achieved.

## 2.7	Research Gap

In view of the above related works, most of the researchers focused mainly on implementing and embedding of sensors to detect the presence of an intruder and sending SMS through GSM module to notify the project owner, but paid less attention to the expensive cost of delivery rate of the sent SMS. The approach adopted in this research is targeted at ensuring the property owner is properly notify and alerted through a phone call in the event of the intrusion detection.

# CHAPTER THREE
# RESEARCH METHODOLOGY

## 3.1 System Design

The intrusion detection system for homes and offices, which uses GSM module and some interface which consist of different units, which include microcontroller unit, sound sensor unit, magnetic switch unit, motion sensor unit with communication unit and notification unit. The sensors are to detect unusual changes in the home and office premises that go beyond typical levels and bounds. The sensors there after transfer their detection data to the ATMega328 microcontroller after identification. The ATMega328 microcontroller analyze and then transform data been sent from the detection of the sensor unit to generate a nearly early detection information. Using the data from sensor detection information, the microcontroller will send commands to the GSM module. Through phone call, the GSM module will then communicate and call the house owner or office owner mobile phone if there is an intrusion.

The system design consists of hardware design and software design, implemented together to achieve this project. Figure 3.1 show the block diagram of the system and Figure 3.2 shows the system flowchart. The system has several interface of hardware components such as the ATMega328 microcontroller, sound sensor, magnetic switch, motion sensor, GSM module (SIM 800L), jumper cable, signal cable and some other electronic and electrical component. The software part consists of code that was compiled on the Arduino IDE and embedded into the microcontroller. The designing and construction of an intrusion detection system for home and office using GSM module for feedback was designed to reduce the crime, robbery and abduction that happened as result of home owner and office owner absence and late delivery of SMS messages.

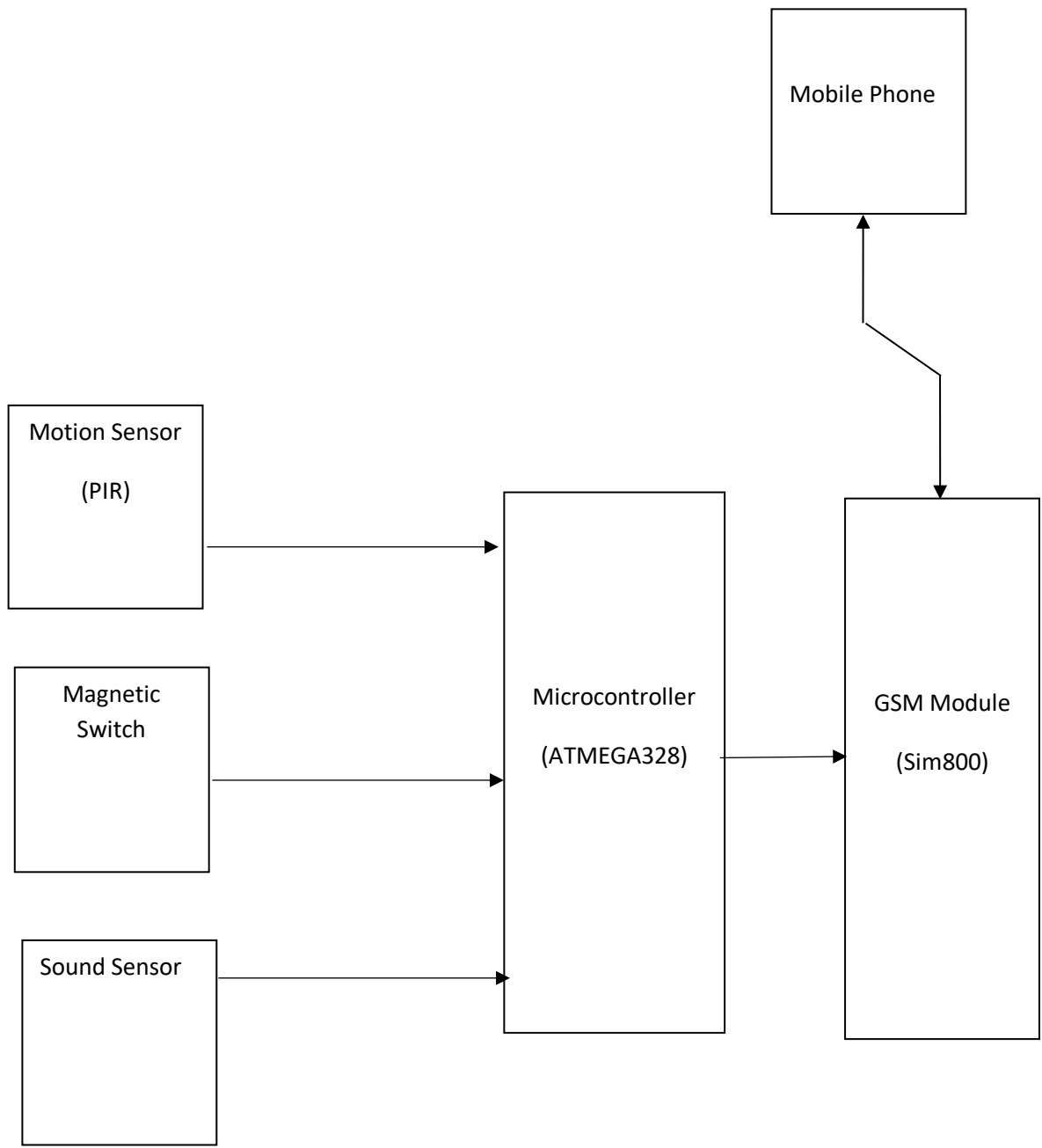

Figure 3.1: Generalized Block Diagram of GSM Based Intrusion Detection System

### 3.2 Passive Infrared Sensor

Passive Infrared Sensor is a low cost, low power and reliable sensor. Therefore, it was felt that a PIR sensor-based security system consisting of the sensor, alighting system and a recording system. The sensor used can detect the presence of intruders within the range of 90 degrees. Upon detection of IR, PIR sensor generates the output in the form of electrical signal. Although the output from the sensor is of few volts, it could be amplified to required voltage using amplifier circuit and could be used for actuating lighting system. The PIR sensor is shown in Figure 3.2.

### 3.3 Sound Sensor

Sound sensor is one of the important sensors implemented in this project, integration of the sound sensor was aimed to enhance the system's ability to detect intrusions accurately by analyzing variations in sound levels within the monitored environment. The sound sensor was implemented into the system architecture by connecting it to the microcontroller (ATMEGA 328). A sound sensor instead consists of a peak detector, an amplifier (LM393, LM386, etc) that is highly sensitive to sound, and an inbuilt capacitive microphone. Sound sensors, having these components, can function correctly. They follow the following process to "hear" sound:

- Sound waves are propagated through air molecules.
- The sound waves are received by the inbuilt capacitive microphone.
- The sound waves are then amplified and digitized for the processing of sound intensity.

Through this process, sound sensors can detect when you clap your hands to switch on the light or when you want to switch them off. The sound sensor is shown in Figure 3.3.

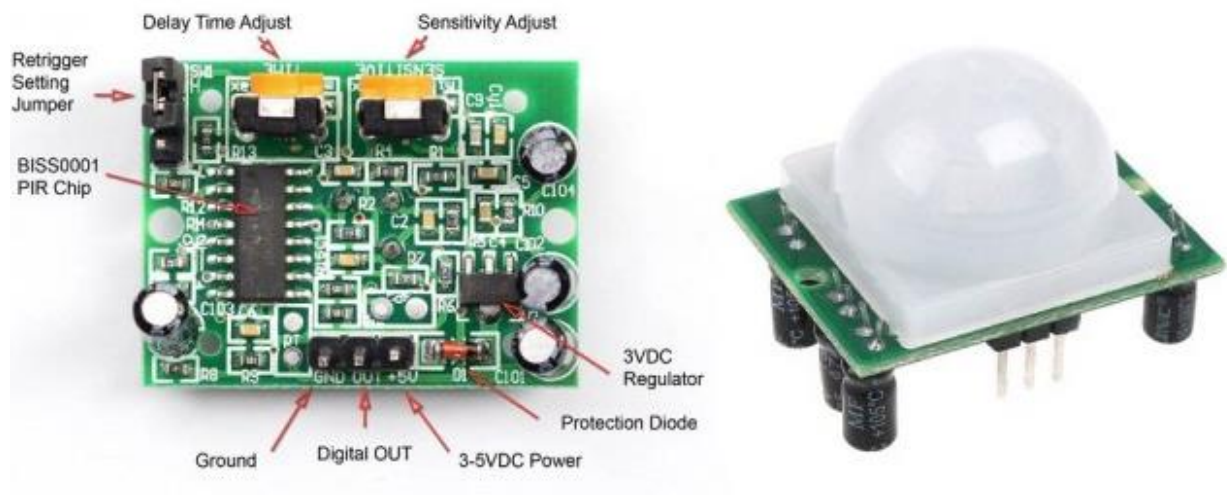

Figure 3.2: Passive Infrared Sensor

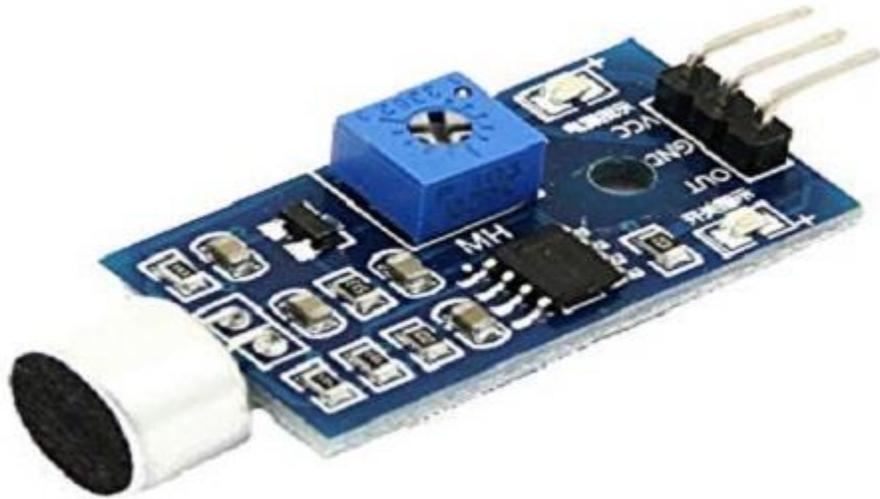

Figure 3.3: Sound Sensor

### 3.4 Magnetic switches

The magnetic switch sensors are integrated with the microprocessor unit, which serves as the central processing unit of the system. Sensor signals are fed into the microprocessor unit input ports, where they are processed in real-time to detect intrusion events. Magnetic switch consist of a magnet, which is installed on a door or window, and a switch unit, which is installed on the frame or jamb. When the door or window is in the closed position, the created circuit is closed; when the door or window is opened, the circuit is open, which causes an alarm to be initiated. A Balanced Magnetic Switch (BMS) employs a magnet in both the switch and magnet units. The switch unit, which contains a magnetic reed switch, a bias magnet, and tamper/supervisory circuitry, is mounted on the stationary part of the door or window unit. The component containing the larger permanent magnet is mounted on the movable part of the door or window, adjacent to the switch unit installed on the frame or jamb when the door is closed. With the door or window closed, the magnets are adjusted to create a magnetic loop, causing the reed switch to experience a magnetic field of essentially zero. When the door or window is opened and the magnetic field is removed, the contacts will separate and trigger an alarm indicating a security breach. The magnetic switch is shown in Figure 3.4.

### 3.5 GSM Module (SIM 800L)

Global System for Mobile communication (GSM) module is a mobile communication modem. GSM is an open and digital cellular technology used for transmitting mobile voice and data services operate at the 850MHz, 900MHz, 1800MHz and 1900MHz frequency bands. A GSM module is employed in this design to enable the fire alarm system place a call to the owner in the event of a fire. This design used SIM800L because it is a very small quad-band GSM module which communicates with any microcontroller through UART serial communication. It only requires 4 wires to communicate with the microcontroller: VCC (5V), RXD (Receive data). TXD (Transmit data) and GND (Ground, 0V). The GSM module is shown in Figure 3.5.

**3.6  Principle of Operation**

The GSM based intrusion detection system for homes and offices was designed to ensure that the presence of intrusion is intelligently detected, promptly notified and interactively managed than what is obtainable with conventional intrusion detection systems. The heart of this design of intrusion detection is a microcontroller ATmega328P, programmed to monitor the outputs of the sensors employed to monitor the presence of an intrusion. To detect the presence of an intruder, the system uses PIR sensors to monitor the presence of any motion and the sound sensor for presence of unauthorized voice and employs magnetic switches to additionally detect unauthorized entry by monitoring the opening or closing of doors or windows, providing an extra layer of security to detect any movement of an intrusion.

When there is presence of intruder, associated with PIR sensor, sound sensor and magnetic switches, is detected the GSM-based intrusion detection system activate the GSM module and gives a phone call to the home owner or office owner by dialing the pre-registered numbers to alert them of the presence of intruder. The circuit diagram illustrating the principle of operation is shown in Figure 3.6

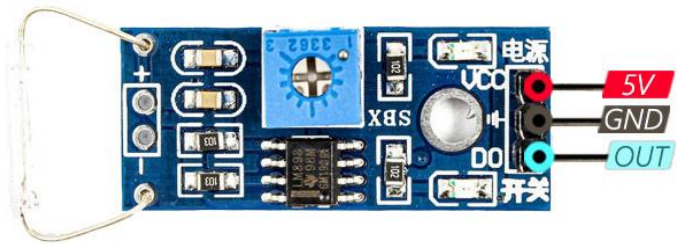

Figure 3.4: Magnetic Switch

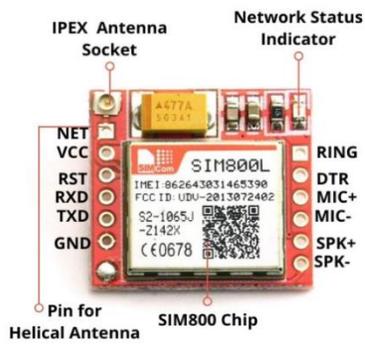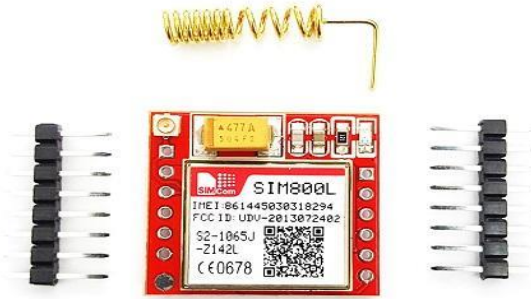

Figure 3.5: SIM800L GSM module

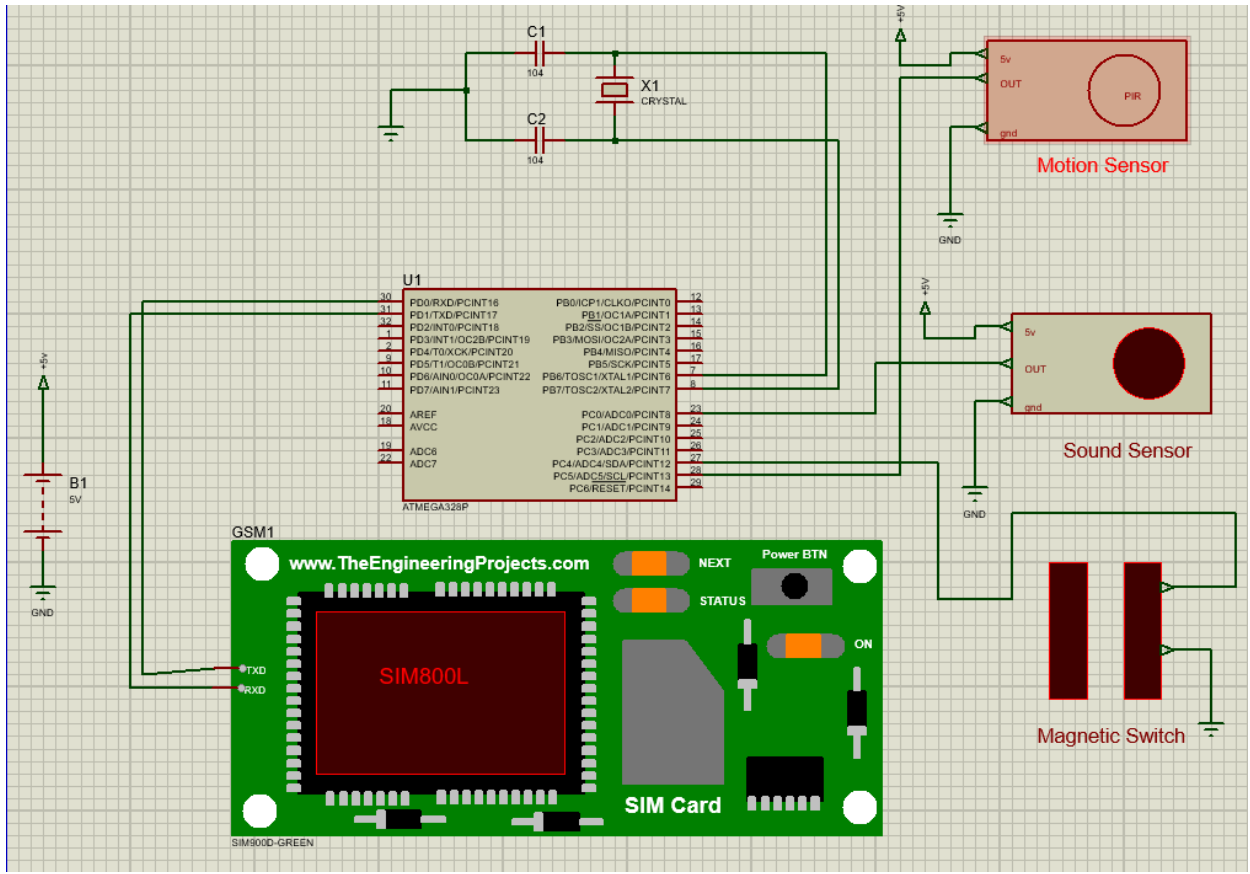

Figure 3.6: Circuit diagram for the GSM-based intrusion detector for homes and offices

# CHAPTER FOUR
# CONSTRUCTION AND TESTING

## 4.1 System Implementation

In this chapter, we delve into the practical aspects of implementing the Multi-Sensor GSM-based intrusion detection system for homes and offices. The implementation phase involves several key steps, including bread boarding, soldering, testing, and casing of the system.

### 4.1.1 Bread Boarding

Building the GSM-based intrusion detection system, bread boarding serves as the initial step used in constructing the hardware components of the intrusion detection system. The breadboard was used to assemble, connect all the components and provide a temporary platform for arranging the ATMega328 microcontroller, sound sensor, magnetic switch, motion sensor, GSM module, and other necessary electronic parts.

### 4.1.2 Soldering

The various circuits and stages of this project were soldered in tandem to meet desired workability of the project and this was done on a printed circuit board as shown in the figure 4.1. The printed circuit board has the power supply stage, which includes; the rectification stage, filtering stage and the voltage regulation stage. It also has the microcontroller stage and the GSM module stage.

### 4.1.3 Testing

Stage by stage testing was done according to the block representation on the breadboard, before soldering of circuit commenced on printed circuit board. The process of testing and implementation involved the use of some test and measuring equipment stated below.

1. **Bench Power Supply**: This was used to supply voltage to the various stages of the circuit during the breadboard test before the power supply in the project was soldered. Also, during the soldering of the project, the power supply was still used to test various stages before they were finally soldered.

2. **Oscilloscope:** The oscilloscope was used to observe the ripples in the power supply waveform and to ensure that all waveforms were correct and their frequencies accurate. The waveform of the oscillation of the crystal oscillator used was monitored to ensure proper oscillation at 16MHz.

3. **Digital Multi-meter:** The digital multi-meter basically measures voltage, resistance, continuity, current, frequency, temperature and transistor hfe. The process of implementation of the design on the board required the measurement of parameters like, voltage, continuity, current and resistance values of the components and in some cases frequency measurement. The digital multimeter was used to check the output of the voltage regulators used in this project.

### 4.1.4 Casing

The third phase of the project construction is the casing of the project. This project was coupled to a plastic casing as shown in the figure 4.1a and 4.1b respectively. The casing material being plastic designed with special perforation and vents and also well labeled to give ecstatic value.

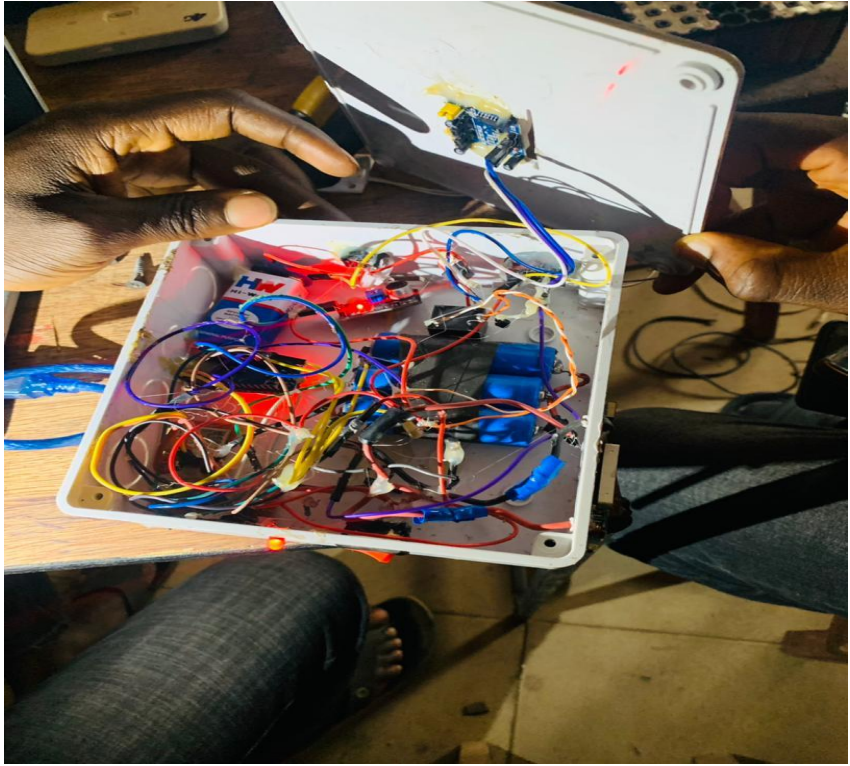

Figure 4.1a: Connection of Components

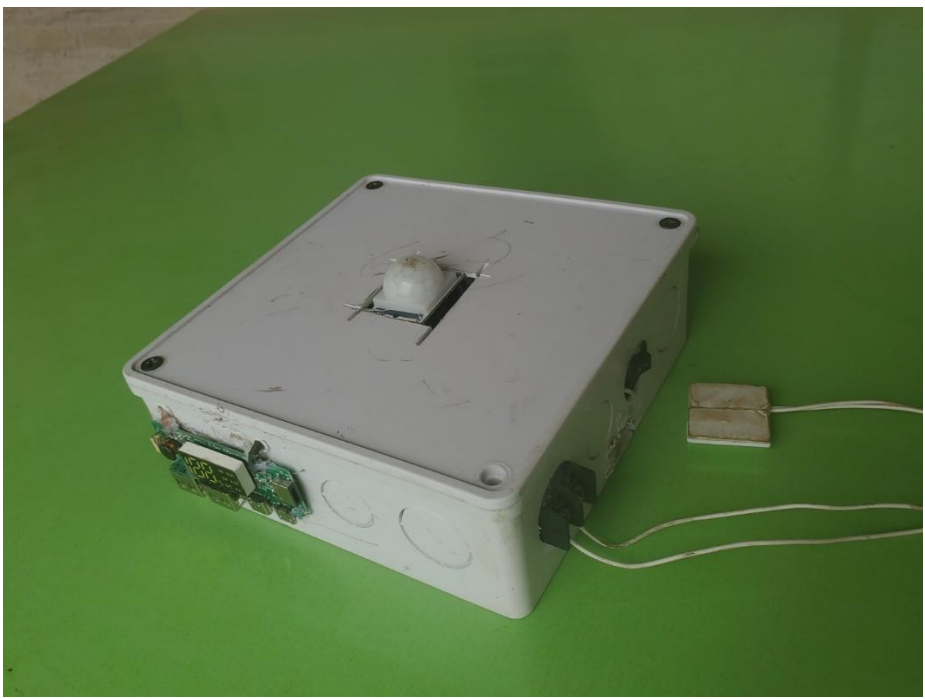

Figure 4.2b: System after Casing

### 4.1.5    Final Testing

After casing the components of the GSM-based intrusion detection system, a final testing phase was conducted to ensure its functionality in real-world conditions. Various intrusion scenarios were simulated to evaluate the sensors' accuracy in detecting unusual activity and triggering the GSM module for phone call alerts. Specifically, the PIR sensor, sound sensor, and magnetic switch were tested to assess their responsiveness to potential intrusions. Once an intrusion was detected, the system promptly initiated a phone call through the GSM module to alert the homeowner or office owner, enabling swift intervention and ensuring enhanced security. Overall, this thorough testing process validated the system's reliability and effectiveness in providing comprehensive intrusion detection and notification capabilities.

### 4.2    Hardware Implementation

The system's operation begins with the magnetic switches installed on the door of the house. These switches are finely tuned to detect any attempts to open the door, whether it's forcefully pushed or opened with the slightest of nudges. When the door is manipulated in any way, the magnetic switch is immediately triggered, sending a signal to the central control unit of the system. Upon detecting this unauthorized entry, the system springs into action, swiftly analyzing the situation. With precision and accuracy, it identifies the breach and promptly alerts the property owner via phone call as seen in Figure, 4.2a and 4.2b respectively. This immediate notification ensures that the property owner is promptly informed of the security breach, allowing them to take swift action to address the situation. Whether it's contacting law enforcement, remotely monitoring the premises, or activating additional security measures, the property owner can respond promptly to protect their assets and ensure the safety of their property. By providing real-time alerts, the system enables proactive security management, minimizing the risk of further intrusion and safeguarding the premises against potential threats.

In addition to the magnetic switches, the system is equipped with a Passive Infrared (PIR) sensor, designed and constructed at a 90-degree angle. This specialized sensor is finely tuned to detect any movement and motion within its designated range. When the PIR sensor detects any activity within its 90-degree field of view, it immediately signals the microprocessor of the system. Upon receiving this signal, the system springs into action, swiftly analyzing the detected movement. The system promptly alerts the property owner via phone call as seen in Figure, 4.3a and 4.3b respectively. This immediate notification ensures that the property owner is promptly informed of the security breach, allowing them to take swift action to address the situation. By combining the capabilities of both the magnetic switches and the PIR sensor, this system provides comprehensive intrusion detection coverage, ensuring the safety and security of the premises.

**4.3      Discussion**

The Multi-Sensor GSM-based intrusion detection system was designed and executed to provide enhanced security for homes and offices. By integrating multi-sensor components with a microcontroller and GSM module, a reliable system capable of detecting potential intrusions in real-time was created. The sensors continuously monitor the surroundings for any unusual activity, such as motion, sound, or unauthorized entry. Once an intrusion is detected, the microcontroller swiftly analyzes the data and triggers the GSM module to alert the homeowner or office owner via phone call. This seamless integration of components ensures prompt notification and enables proactive responses to security threats, ultimately enhancing overall safety and peace of mind for users.

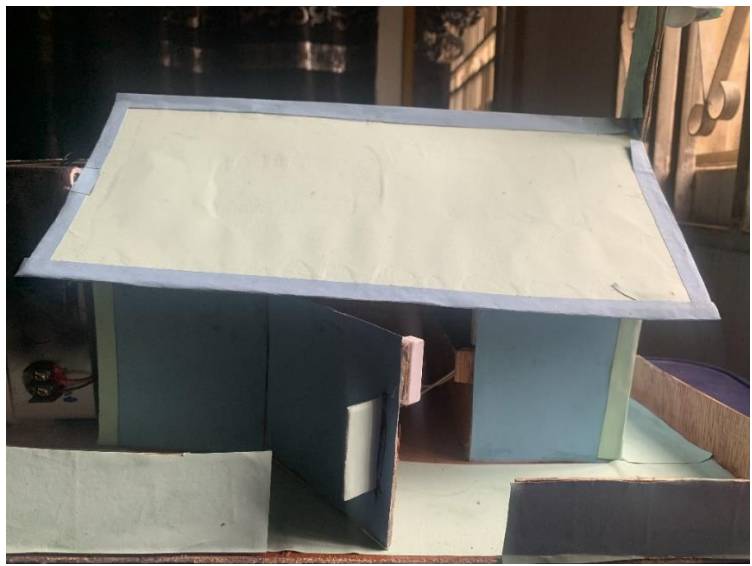

Figure 4.2a: Door intrusion detection diagram

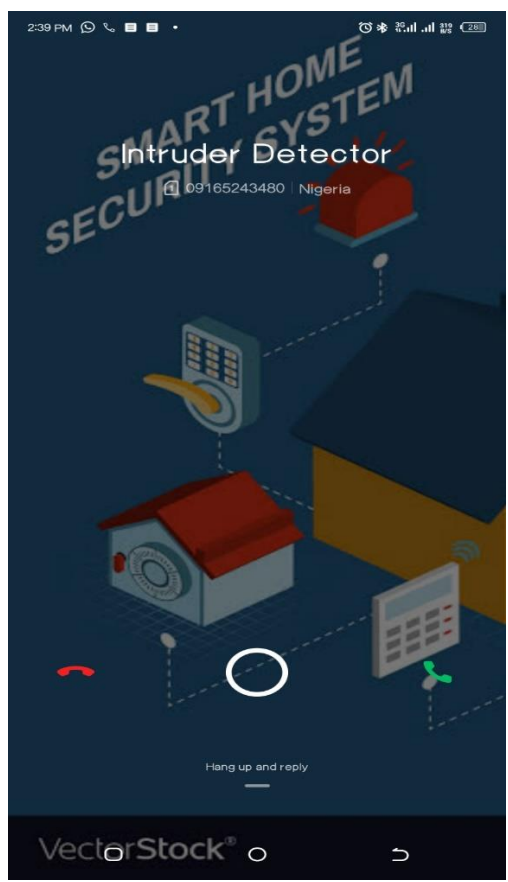

Figure 4.2b: Call from the system after detecting an intruder

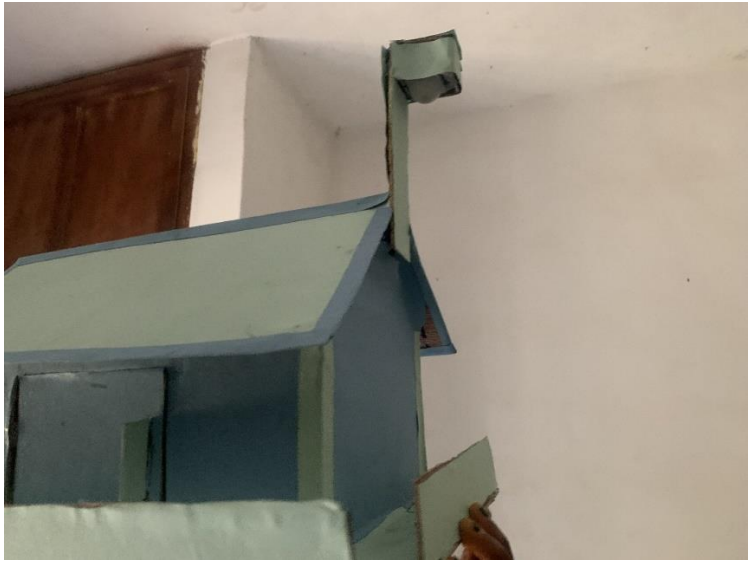

Figure 4.3a: PIR sensor detection coverage diagram..

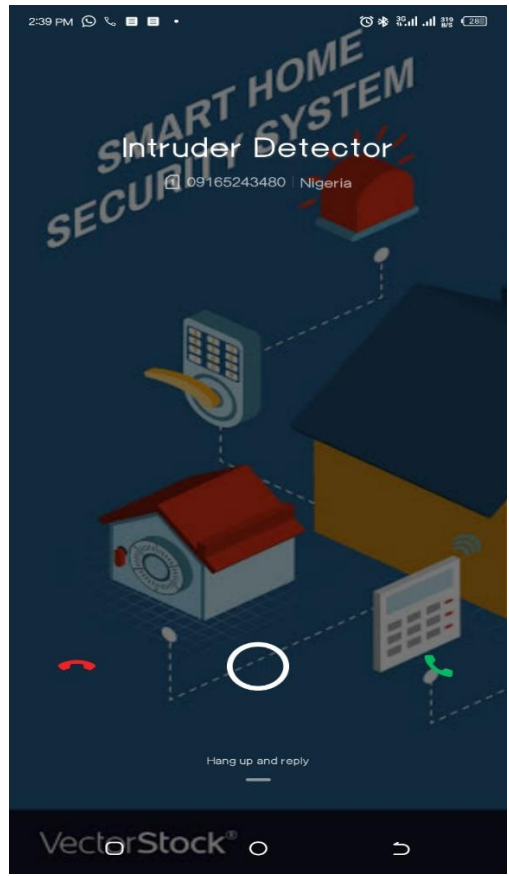

Figure 4.3b: Call from the system after detecting an intruder

# CHAPTER FIVE

# CONCLUSION AND RECOMMENDATION

## 5.1 Conclusion

This research work designed and implemented a multi-sensor intrusion detection system using three detecting sensors; PIR sensor, sound sensor and magnetic switches; a phone call gateway via GMS is used for feedback. The system indicated the occurrence of intruder at its early stage by calling the property owner whenever one of the input sensors detect any unusual activity, such as motion, sound, or unauthorized entry. The implementation of the system involved breadboarding, soldering, testing, and casing. Each component was carefully assembled and tested to ensure functionality and reliability. Through rigorous testing, the system demonstrated its capability to detect intrusions effectively and provide prompt phone call alerts to the property owner.

## 5.2 Recommendation

Nowadays, more houses and offices are utilizing smart security systems to prevent unauthorized access. But the multi-sensor intrusion detection system implemented in this project won't operate and function well when the battery runs out. Hence, to improve it, researchers in the future can use solar power or a rechargeable battery to power it even when the battery is nearly empty. In this manner, the system will constantly be vigilant and prepared to keep property safe from intruders.

Additionally, home and office owners should be educated about the importance of security measures and the functionality of intrusion detection systems to encourage wider adoption and promote proactive security practices. Furthermore, integration with existing security infrastructure such as surveillance cameras, and access control systems should be prioritized. This ensures a clear picture of intruders is captured and facilitates comprehensive security coverage, leading to better incident response.

# APPENDIX

```cpp
#include <SoftwareSerial.h>
SoftwareSerial gsm(11,10);
void setup() {
   Serial.begin(9600);
   gsm.begin(9600);
 Serial.println("initializing...");
  delay(15000);
 pinMode(8,INPUT_PULLUP);
  pinMode(7,INPUT_PULLUP);
  pinMode(9,OUTPUT);
}

void loop() {
int val = analogRead(A0);
 Serial.println(val);
 delay(500);
 if (val>800){
Serial.println("noise detected");
digitalWrite(9,1);
delay(1000);
digitalWrite(9,0);
delay(1000);
// gsm.print("ATD+2347048850497;\r\n");
//   Serial.println(gsm.readString());
```

```
//      delay(2000);
 }

else{
  Serial.println("Safe mode");
}
Serial.println("output");
if(digitalRead(7)==0){
Serial.println("PIR");
Serial.println(digitalRead(7));
digitalWrite(9,1);
delay(1000);
digitalWrite(9,0);
delay(1000);
 gsm.print("ATD+2347048850497;\r\n");
  Serial.println(gsm.readString());
       delay(2000);
}

if(digitalRead(8)==1){
Serial.println("Magnetic");
Serial.println(digitalRead(7));
digitalWrite(9,1);
delay(1000);
digitalWrite(9,0);
```

```
  delay(1000);
 gsm.print("ATD+2347048850497;\r\n");
   Serial.println(gsm.readString());
       delay(2000);
       message();
}
}

void message(){
gsm.println("AT+CMGF=1");
    delay(1000);
    gsm.println("AT+CMGS=\"+2347048850497\"\r");
    delay(1000);
    gsm.println("ALERT!!\n Intruder detected!!!");
    delay(1000);
    gsm.write((char)26);
    delay(1000);
    Serial.println(gsm.readString());
    delay(1000);
}
```